# CHARACTERISTICS OF POST PARTUM PERINEUM WOUNDS WITH INFRA RED THERAPY

**Bina Melvia Girsang[1], Eqlima Elvira[2]**
Nursing Faculty, Universitas Sumatera Utara
Department of Maternity and Child Nursing [1], Department of Surgical Medical Nursing [2]
Correspondence email : binamelvia@usu.ac.id

**Abstract**
Perineal wounds other than their localization in humid feminine areas and can pose a risk of infection. This quantitative study aims to see a description of the characteristics of post partum maternal perineal wounds with infrared therapy twice a day for three consecutive days with a descriptive analytic analysis design. Sample selection technique with purposive sampling in post partum mothers with spontaneous parturition, 1-2 degrees of perineal wounds in the amount of 20 people. Perineal wound characteristics were assessed and observed using the Southampton instrument with the results of the instrument trial r = 0.99 (Karl Pearson correlation coefficient), and the reliability value of 0.99 (brown spearman). The results of this study showed that degeneration of wounds was found only until the first day of post-intervention (1.95 ± 0.22), poor wound regeneration occurred until the second day of pre-intervention (1.95 ± 0.22), and wound regeneration was being encountered changes from the first day to the second day pre-intervention (1.05 ± 0.22). These results indicate that infrared therapy does not show significant changes if only done for one day.

Key Words: localization, parturition, regeneration

## Introduction

Perineal infection can occur due to the location of the perineum that is moist so that it becomes a medium where bacteria develop. The incidence of infection that occurs in the perineal wound can spread to the area of the birth canal or the urinary tract. An infection condition in the perineal wound will slow the wound healing process because it can add damage to the supporting tissues in the skin. This condition will aggravate the degree of perineal injury and treatment[1]. The period where after the mother gave birth to the placenta at the third stage, then the mother will enter a recovery period called the post-natal period. At this time the post partum mother will experience a recovery process both physically and psychologically. In this postpartum period, the mother will undergo a new role as a mother and will experience a process of uterine involution[2]. Mothers experience some discomfort after giving birth even though they are considered to be common discomforts during the puerperium after birth pain, perineal pain, fatigue, constipation, breast swelling, lactation suppression, headache, back pain, can cause physical[2], discomfort, psychological stress and poor quality of life of the mother[3].

Some treatments to relieve perineal pain and promote wound healing. Pharmacological and non-pharmacological methods are used to treat this discomfort. Pharmacological pain relief methods include non-steroidal anti-inflammatory drugs, oral analgesics, local anesthetics and opioids. But this method is associated with serious side effects such as constipation, gastric irritation, the course of drugs for breast milk, and cases of prolonged bleeding[4].

With regard to non-pharmacological methods, common practice is the use of ice packs, and the application of heat. Ice packs during the first 24 hours postpartum are a traditional method used for immediate pain relief symptoms because it sedates the perineum, but this help is generally short-lived, and there is no evidence of long-term benefits. After 24 hours, heat is recommended because it increases circulation to the region. The form of heat used is the application of sitz bath therapy or infrared lamps. It helps reduce perineal edema, to avoid hematoma formation, to eliminate discomfort, to encourage wound healing by cleaning the perineum and anus, and reduce inflammation[5],[6]. In the results of the study mentioned that infra red therapy is a method of treatment of the perineum by utilizing the effect of 245 volts of infrared light, at a distance of 45-50 cm with a duration of 10-15 minutes will have an impact to provide comfort and reduce pain in perineal wounds[7].

Nurses play an important role in the treatment of perineal wounds. Prevention of infection needs to be done given that there is always a danger that bacteria can spread from the rectal area to the vagina. Some relevant interventions have been carried out in perineal care and further research is still needed in order to be a procedure that can minimize the incidence of infection, besides considering continuity of procedures that should be easy to do and economically valuable so as to accelerate the process of healing of the perineal wound and the mother can be more prosperous in undergoing childbirth[8]. Based on these considerations the researchers examined infrared therapy performed at Sundari General Hospital, Medan, North Sumatra, Indonesia. This research is a quantitative study that aims to see a description of the characteristics of post partum maternal perineal wounds with infra red therapy.

## Method

This research is a descriptive analytic study to assess the characteristics of perineal wounds in post partum mothers with infra red therapy. The sampling technique was done by purposive sampling, selected 20 postnatal mothers with spontaneous (normal) labor, 1-2 degree perineal injuries. Perineal wound observation was carried out using the Southampton instrument with the results of the instrument trial r = 0.99 (Karl Pearson correlation coefficient), and the reliability value of 0.99 (brown spearmen). Infra red therapy is carried out for three days with a frequency of twice a day in the morning and evening after the post partum mother has finished bathing. Administrative considerations and ethical approval results have been permitted to conduct this study with an ethics approval letter issued by the Faculty of Nursing, University of North Sumatra with letter number 1724 / IV / SP / 2019. The research subject was given informed consent after an explanation of the purpose of the study was made where the confidentiality of the research subjects was guaranteed by the researcher

In data management, all complete questionnaires are captured and administered in the Statistical Package for Social Sciences (SPSS) version 19. database management system. Data is then cleaned to ensure that only valid responses to questions are present in the database. Logic checks are also carried out. To make data more meaningful, frequency or percentage tables, descriptive statistics, and inferential statistics are used to analyze and present data.

## Results

In this study a univariate analysis was performed on the characteristics of postpartum maternal wounds with infrared therapy. Tabulation of the frequency distribution of perineal wound healing can be seen in the table below

Table 1
Frequency Distribution of Postpartum Mothers With the Healing Process of Perineum Wounds Through Infra Red Therapy

| No | Wound Healing Interpretation | Therapy Infra Red | | | | | | | | | | | |
|---|---|---|---|---|---|---|---|---|---|---|---|---|---|
| | | Day-1 Pre | | Day-1 Post | | Day-2 Pre | | Day-2 Post | | Day-3 Pre | | Day-3 Post | |
| | | f | % | f | % | f | % | f | % | f | % | f | % |
| 1 | Wound Degeneration | 1 | 5 | 1 | 5 | | | | | | | | |
| 2 | Poor wound regeneration | 4 | 20 | 3 | 15 | 1 | 5 | | | | | | |
| 3 | Moderate wound regeneration | 15 | 75 | 16 | 80 | 19 | 95 | 20 | 100 | 20 | 100 | 20 | 100 |
| 4 | Good wound regeneration | | | | | | | | | | | | |

In infra red therapy on the first day before the intervention, there were 4 (20%) postpartum mothers had poor wound regeneration category and 15 (75%) postpartum mothers with moderate wound regeneration category, after twice intervention there was a decrease in the results of the wound regeneration category the less good is 3 (15%) postpartum mothers. On the second day only 1 (5%) postpartum mothers with a category of poorly regenerated wounds before intervention, and after the intervention found that 20 (100%) mothers with a category of moderate wound regeneration. The results of the analysis on the third day showed the same results after the intervention on the second day, where before and after the intervention of infra red therapy the wound healing process category did not change (100%) with the category of moderate wound regeneration. The average wound healing process with infrared therapy is explained in the table below.

Table 2
Average Characteristics of Post Partum Perineum Wounds with Infra Red Therapy Twice a Day for Three Days

| Interpretation Characteristics of Perineal Wounds | Day-1 | | | | Day-2 | | | | Day-3 | | | |
| --- | --- | --- | --- | --- | --- | --- | --- | --- | --- | --- | --- | --- |
| | Pre | | Post | | Pre | | Post | | Pre | | Post | |
| | Mean | SD | Mean | SD | Mean | SD | Mean | SD | Mean | SD | Mean | SD |
| Wound Degeneration | 1,95 | 0,22 | 1,95 | 0,22 | 2,00 | 0,00 | 2,00 | 0,00 | 2,00 | 0,00 | 2,00 | 0,00 |
| Poor wound regeneration | 1,80 | 0,410 | 1,85 | 0,37 | 1,95 | 0,22 | 2,00 | 0,00 | 2,00 | 0,00 | 2,00 | 0,00 |
| Moderate wound regeneration | 1,10 | 0,31 | 1,10 | 0,31 | 1,05 | 0,22 | 1,00 | 0,00 | 1,00 | 0,00 | 1,00 | 0,00 |
| Good wound regeneration | 2,00 | 0,00 | 2,00 | 0,00 | 2,00 | 0,00 | 2,00 | 0,00 | 2,00 | 0,00 | 2,00 | 0,00 |

Based on the above table, it is explained that the description of the characteristics of wounds on the first day showed no difference. On the first day degenerated perineal wounds showed no changes before and after infra red intervention was given twice a day (1.95 ± 0.22) in one post partum mother. But on the second and third days there were no more post partum maternal perineal injuries that degenerated. Likewise, good wound regeneration did not occur in infra red therapy for three consecutive days (2.00 ± 0.00). Changes in the condition of perineal wounds undergo poorly regenerated wounds on the first day also not found again after the intervention of infrared therapy on the second and third day (2.00 ± 0.00). The condition of repair of perineal wounds toward the regeneration of wounds that are happening is more dominant, starting from the first day to the second day pre infra red therapy intervention (1.05 ± 0.22), and then constant at the post intervention on the second day until post therapeutic intervention infra red day three (1.00 ± 0.00). Changes in wound characteristics can be seen clearly in the graph below.

Graph 1
Distribution of Average Characteristics of Postpartum Maternal Injuries With Infra Red Therapy Twice a Day for Three Consecutive Days

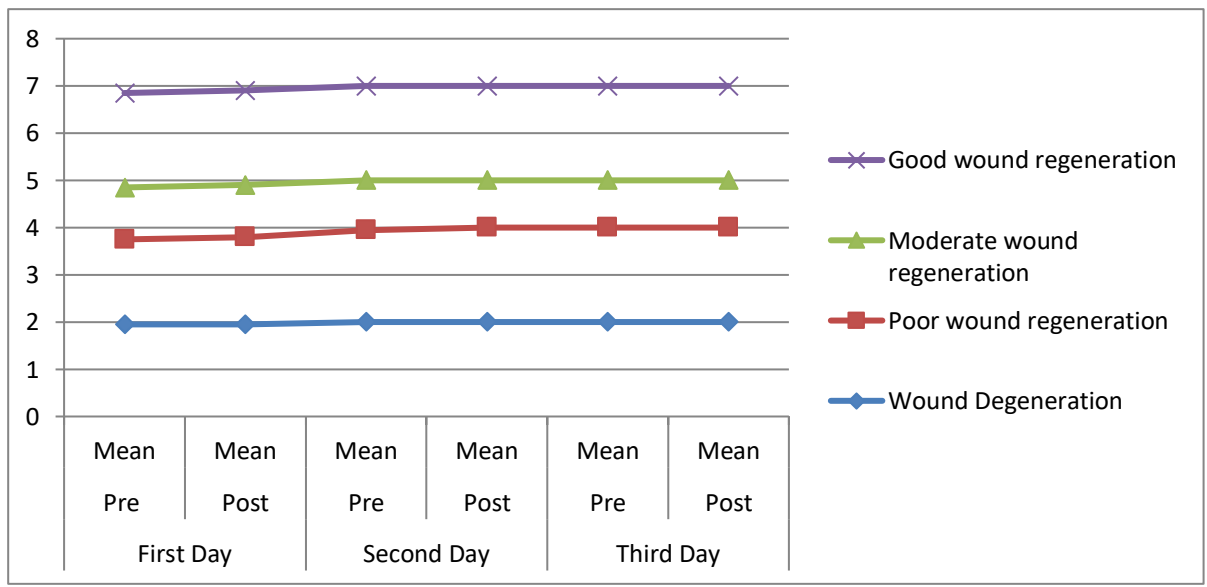

## Discussion

The use of the effects of infrared therapy is proven to be able to improve the state of the wound. The results of research on diabetes wounds given infra-red therapy for 30 minutes of use can improve circulation in the injured foot[9]. Using measurement sievers instruments, Model 280, Nitric Oxide Detectors (the authors of unpublished data) found that local and systemic nitric oxide can increase circulation even though the effect is still visible even though infrared therapy has been knocked out[10]. Some literature states that the wound healing phases there are 3 phases, namely, the inflammatory phase, the proliferation phase, and the maturation phase[11].

The process of repair of perineal skin tissue is influenced by several factors, including the location of the wound, the severity, and the degree of injury[12]. The process of wound regeneration is characterized by the presence of new blood vessels as a result of tissue reconstruction and occurs within 3-24 days, characterized by changes in color characteristics, or texture of the perineal wound[13]. In this study the assessment of wound regeneration conditions were divided into 4 categories namely wound degeneration, poorly wound regeneration, moderate wound regeneration, and good wound regeneration. Good wound regeneration did not occur in infra red therapy for three consecutive days (2.00 ± 0.00). Changes in the condition of perineal wounds undergo poorly regenerated wounds on the first day also not found again after the intervention of infrared therapy on the second and third day (2.00 ± 0.00). The condition of repair of perineal wounds toward the regeneration of wounds that is happening is more dominant, starting from the first day to the second day pre infra red therapy intervention (1.05 ± 0.22), and then constant at the post intervention on the second day until post therapeutic intervention infra red day three (1.00 ± 0.00).

The process of healing perineal wounds can also be inhibited by several things, namely, the presence of infection, the moisture balance of the perineum skin. Based on research presented by Istiqomah in 2014[14], that the episiotomy wound healing process given infra red therapy with a wavelength of 770nm-106nm with a distance of 30 cm and for 15 minutes, the value of the Mann Whitney 41.5 statistical results and the value of P value 0.003 <0, 05, where the wound healing process occurs for 7 days, compared with those not treated experiencing healing in 9 days. The first day on this study, before the intervention, there were 4 (20%) postpartum mothers had poor wound regeneration category and 15 (75%) postpartum mothers with moderate wound regeneration category, after twice intervention there was a decrease in the results of the wound regeneration category the less good is 3 (15%) postpartum mothers. On the second day only 1 (5%) postpartum mothers with a category of poorly regenerated wounds before intervention, and after the intervention found that 20 (100%) mothers with a category of moderate wound regeneration.

The effect of perineal wound healing with infra red therapy on postpartum mothers is a unique treatment, wherein this therapy uses infrared light exposure on perineal wounds or on the diseased part. This research is felt to be very important because nurses can control the condition of repair of perineal wounds so that they can plan appropriate treatment actions on perineal wounds. The correct treatment method and can heal wounds in a short time is very important[15]. The use of non-pharmacological measures such as infrared therapy can be included with other contemporary therapies. To equip nurses to provide holistic care to their clients, the nursing practice must be closed with several types of non-pharmacological measures such as infrared therapy for wound healing. Thus the nurse can be guided in developing the right attitude and skills to treat patients with perineal wounds.

## Conclusion

Degeneration of wounds was found only until the first day of post-intervention (1.95 ± 0.22), poor wound regeneration occurred until the second day of pre-intervention (1.95 ± 0.22), and wound regeneration was being encountered changes from the first day to the second day pre-intervention (1.05 ± 0.22). These results indicate that infrared therapy does not show significant changes if only done for one day.

## Acknowledgements

The researchers thank the Universitas Sumatera Utara for giving ethical permission and funding in this study. Special thanks to the postpartum mothers who participated in this study.